\documentclass[manuscript]{aastex}

\usepackage{subfigure}
\graphicspath{{figures/}}

\shorttitle{Multi-Scale Morphological Analysis of SDSS DR5 Survey}
\shortauthors{Yongfeng et al.}

\begin{document}


\title{Multi-Scale Morphological Analysis of SDSS DR5 Survey \\ Using the Metric Space
Technique}


\author{Yongfeng Wu, David J. Batuski}
\affil{Department of Physics and Astronomy, University of Maine, Orono, ME 04469, USA}

\author{Andre Khalil}\affil{Department of Mathematics \& Statistics \\ and
  Institute for Molecular Biophysics, University of Maine, Orono, ME 04469, USA}
\email{yongfeng.wu@umit.maine.edu}


\begin{abstract}
Following novel development and adaptation of the Metric Space Technique (MST), a
multi-scale morphological analysis of the Sloan Digital Sky Survey (SDSS) Data Release 5 (DR5) was
performed. The technique was adapted to perform a space-scale morphological analysis by filtering
the galaxy point distributions with a smoothing Gaussian function, thus giving quantitative
structural information on all size scales between 5 and 250 Mpc. The analysis was performed on a
dozen slices of a volume of space containing many newly measured galaxies from the SDSS DR5 survey.
Using the MST, observational data were compared to galaxy samples taken from \emph{N}-body simulations with
current best estimates of cosmological parameters and from random catalogs. By using the maximal
ranking method among MST output functions we also develop a way to quantify the overall similarity
of the observed samples with the simulated samples.

\end{abstract}





\section{INTRODUCTION}
From redshift surveys such as the Sloan Digital Sky Survey (SDSS; \citealt{york00}) and the
Two-Micron All Sky Survey (2MASS; \citealt{skr00}), the local universe shows intricate patterns with
clusters, filaments, bubbles, sheet-like structures and the so-called voids. For a review of the
structural analysis of the universe, see \citet{wei05}. At the same time, Lambda Cold Dark Matter $\lambda$CDM models have been developed, see \citet{gill04} and \citet{dol08}. Several simulations have
been created, such as the Millennium Simulation \citep{spr05} done by \citet{cro05} and another
\emph{N}-body simulation by \citet{ber06}. These models describe a universe that consists mainly of dark
energy and dark matter and calculate the evolution of the universe from a short time after the big
bang to the present time. Work has been done to verify the similarity between the real universe and
simulated universe \citep{spr05,ber06} and they agree well, based on the comparative techniques used
in these studies.

To supplement the widely used correlation function and power spectrum,
alternatives have been proposed to quantify structure in the galaxy distribution, such as the genus curve \citep{zel82a},
percolation statistics \citep{zel82a,sha83,sah97}, Rhombic Cell analysis
\citep{wu04}, void probability functions \citep{white79}, high-order
correlation function \citep{peeb80}, and multi-fractal measures
\citep{saar07}. However, all of these consider a single map as a space. Here we generalize the Metric Space Technique (MST), a tool used to analyze and
classify astrophysical maps \citep{ada92}, to perform a multi-scale analysis. Key facets of the MST approach are consideration of any given map as an element in the space of all such maps and definitions of a distance function to make the spaces of all maps into a topological space. Moreover, the other methods focus on summary statistics that convey little of the geometric and topological properties of the galaxy distribution. The MST method gives desired quantitative summary statistics of the difference between maps. However, a primary benefit of our method is that the output functions, such as filamentation, number of components,  density, volume and pixels, are straightforward and simple to understand and particularly useful in maps comparisons. Finally, MST is based on the use of threshold values, which will make sure that we can unambiguously define a space on the map with an interesting topology \citep{ada92}.

The MST allows an objective and quantitative comparison of any two images. All such images are considered to be elements of a metric space, where, instead of comparing images on a pixel-to-pixel basis, the comparison is made by considering the metric distance between two images' output functions. The MST
was first used to analyze Galactic molecular cloud data \citep{ada94,wis94}. Several mathematical
and technical improvements to the technique were presented in \citet{kha04a} (for more details, see
\citet{kha04b}, where the updated formalism was used to analyze Galactic atomic hydrogen gas
regions from the Canadian Galactic Plane Survey \citep{tay03}. For both studies the output functions
were applied to two-dimensional gray-scale images which described a smooth density field. But as
originally suggested by \citet{ada92}, one can choose to smooth point distribution data (e.g.,
galaxy distribution) in order to obtain gray-scale data from which the output functions can be
calculated. The first application to point distribution is done by \citet{wu08}. Importantly,
however, the smoothing level becomes critical in creating the density field \citep{don88,sil81}.
Efforts have concentrated on determining the best density estimate from optimal
smoothing length \citep{cole95,vic05}. 
In this paper, we consider a wide range of smoothing levels for multi-scale filtering \citep{kha06}.
By varying the size of the smoothing function over a range of scales, exactly like the wavelet
transform, a complete multi-scale description of galaxy distributions in metric space becomes
possible.

The goal of this paper is therefore to use the multi-scale MST to quantify morphological differences between the SDSS
observational data and two sets of simulation sample data and then illustrate the use of those differences to understand degrees and types of structure in the galaxy
distribution. Here the
point (galaxy) distribution data was filtered by a smoothing function over a continuous range of
scales. Using this novel approach, the MST not only informs us, quantitatively, about the structure
information of the universe and which mock sample most resembles the observational data, but also
how the information and resemblance vary over size scales.

\section{The Multi-Scale Metric Space Technique}

The formalism has been developed as a form description tool with the aim of comparing any two
different astrophysical maps. In previous studies, any given image would always be compared to a
uniform image where all pixels have the same value \citep{ada94, wis94, kha04a, kha04b}. In this way,
two images were separately compared to a uniform image, giving information on ``how far'' (in the
metric sense) both fall from uniformity, thus quantifying the complexity of each of the maps. This
approach will be used here, but additionally however, the observational data from the SDSS will also
be directly compared to the mock sample data, giving information on how far each mock sample is from
the observations.

\subsection{Output Functions}
\label{sect_output}

Instead of comparing the smoothed maps on a pixel-to-pixel basis, information is extracted from the
maps in the form of {\sl output functions}. An output function is a one-dimensional function
representing a profile of some meaningful physical quantity. Its independent variable is the pixel
value (intensity), called the {\sl threshold value} $\Sigma$, and $\sigma$ denotes a smoothed galaxy
distribution image.

\subsubsection{Distributions of Density and Volume}

The density output function characterizes the fraction $m$ of material at densities higher than the reference threshold
value
$\Sigma$:
\begin{equation}
  m(\sigma;\Sigma) = {\int \sigma(x) \Theta [\sigma(x) - \Sigma] d^2x \over{ \int \sigma(x) d^2x}}
  \label{density}
\end{equation}
where $\Theta$ is a step function and the integrals are taken over a bounded domain, from the
minimal threshold values. This function measures the amount of material occurring at a given
density, reflecting how much material occupies a fixed projected volume. The distribution of density
can be useful to characterize the condensation of material. This is useful in cosmology because
theoretical considerations suggest that galaxies form in the highest density regions.

The distribution of volume characterizes the amount of space occupied by material at a fixed density
level. The distribution of projected volume\footnote{Note that since this study only deals with two-dimensional
images, even though the term {\sl volume} is used, it is the actual distribution {\sl area} (a
projected volume) that is considered.} characterizes the volume fraction $v$ of material at
densities higher than the reference threshold value $\Sigma$:
\begin{equation}
  v(\sigma;\Sigma) = {\int \Theta [\sigma(x) - \Sigma] d^2x \over{ \int d^2x}}
\label{volume}
\end{equation}
This is an important parameter when considering how galaxies are distributed, since the volume
output function will quantify the space filled by galaxies.

\subsubsection{Distribution of Pixels}

The number of pixels representing various numbers of data points is counted in a histogram:
\begin {equation}
  j(\sigma;\Sigma) =\sharp{\Theta [\sigma(x) - \Sigma]}
  \label{pixels}
\end {equation}
where $\Theta$ is a step function and $\sharp$ is the number of elements of the set. This function
is also of interest to be applied to the galaxy distribution, since different spatially distributed
populations can create different histogram shapes.

\subsubsection{Distribution of Topological Components and Filaments}

A topological component is a set of connected pixels in a smoothed map for a fixed threshold value.
We use the notation $n(\sigma; \Sigma)$ to denote the distribution of components (the number of
components as a function of the threshold value). The distribution of components can measure the
connections of material, making it useful to indicate the interaction among galaxies.

Each component can be associated with a filament index, $F$, which characterizes the filamentary
structure of the component. $F$ is defined in the following way:
\begin{eqnarray}
  F = {\pi D^2 \over {4 A}},
\end{eqnarray}
where $D$ and $A$ are the longest straight line between any two parts in the component and the area
of the component, respectively. From this definition one can see that a thin or elongated component
will have a higher filament index than a more circular component, for which $F$ will be close to 1.
It is interesting in cosmology because a thin component is generally the boundary of a void or a
string of galaxies, and the more circular component is generally a cluster. We are interested in the
distribution of the filament indices as a function of threshold value
\begin{eqnarray}
  f(\sigma;\Sigma) = {1\over{ n(\sigma; \Sigma)}}\sum_j F_j,
\end{eqnarray}
where $j = 1, 2,\ldots n(\sigma; \Sigma)$. As originally mentioned in \citet{kha04a} this definition of
the filament index has a fault in that it cannot characterize adequately the filamentary structure
of non-convex objects. A new definition of the filament index is given by
\begin{eqnarray}
  F = {P D \over {4 A}},
  \label{filament}
\end{eqnarray}
where $P$ is the perimeter of the component. The details of the justification for introducing this
new definition are given in the Appendix.

\subsection{The Euclidean Metric, Coordinates, and Maximal Ranking}
\label{sect_Euclidian}

For any functions $f$ and $g$, the Euclidean metric $d_E$ is defined as
\begin{equation}
  d_E(f,g)=\bigg(\int|f(x) - g(x)|^p dx \bigg) ^{1/p},
  \label{Euclidian1}
\end{equation}
where, for this study, $p=2$. If we want to compare a specific output function in two of our maps,
we use the following equation:
\begin{equation}
  d_K(\sigma_{A},\sigma_{B})=\bigg(\sum|K(\sigma_{A};\Sigma) - K(\sigma_{B};\Sigma)|^2 \bigg) ^{1/2}.
  \label{Euclidian}
\end{equation}
Here $\Sigma$ is the threshold value, $K$ is a specific output function and $\sigma_{A}$ and $\sigma_{B}$
are maps. Since $\Sigma$ is discrete in our analysis, we approximate Equation (\ref{Euclidian1}) with a
summation.

In order to obtain the distance between the output functions of the images under study, in this
paper we apply this method in two ways. One way is that the observed images are compared to uniform
images, giving us information on ``how far'' (in the metric sense) the observation fall from
uniformity, thus giving quantitative information on the complexity of observed images. Another way
is that all mock images are compared to observed images, thus, each coordinate gives quantitative
information as to ``how far'' the mock image is from observed data sets. Clearly, the larger each
coordinate is, the ``farther'' the mock image under study is from the observational data.
Coordinates are calculated for each of the output functions, for each of the mock sample data sets,
and for each size scale considered. Following the ranking procedure introduced in \citet{kha04a},
once the coordinates for all output functions are calculated, each coordinate is divided by the
maximal coordinate (out of all mock sample coordinates for a particular output function). These
normalized coordinates are then added to each other for each output function to yield an overall
distance value. For each size scale, this distance value quantifies the difference between each mock
sample and the observational data.

\subsection{Gaussian Filtering}

The two-dimensional Gaussian smoothing function is defined by
\begin{eqnarray}
 G(x,y)=\exp(-|\textbf{x}|^{2}/2)
\label{gauss}
\end{eqnarray}
where $|\textbf{x}|=\sqrt{x^{2}+y^{2}}$. In full analogy with the continuous wavelet transform
\citep{kha06}, Gaussian filtering can be described by
\begin{eqnarray}
 T_{G}[f](\textbf{b},a)=\frac{1}{a^{2}}\int f(\textbf{x})\cdot G(\frac{\textbf{x}-\textbf{b}}{a})d^{2}\textbf{x}
 \label{gauss_filter}
\end{eqnarray}
where $f$ is a two-dimensional function representing the image under study, $G(\textbf{x})$ is the
Gaussian function (Equation (\ref{gauss})), which can also be defined as a wavelet. $a$ is the scale
parameter, and $\textbf{b}$ is a position vector. Thus, the convolution between the point distribution
images under study and the Gaussian filter at several different values of the scale parameter $a$
yields the continuous gray-scale images from which the output functions and then the coordinates can
be calculated.

\section{DATA}
The observational galaxy sample was taken from the SDSS DR 5 \citep{jen07}. DR 5
includes five-band photometric data for 217 million objects selected over 8000 deg$^{2}$, and
1,048,960 spectra of galaxies, quasars, and stars selected from 5713 deg$^{2}$ of that imaging data.

This sample of galaxies is approximately complete down to an apparent $r$-band Petrosian magnitude
limit of 17.77, with absolute magnitudes \emph{k}-corrected \citep{bla05}. In order to limit the effects of
incompleteness on our group identification, we restrict our sample to regions of the sky where the
completeness (the ratio of obtained redshifts to spectroscopic targets) is greater than 90\%, and $r$-band
magnitude limit is 17.5 (this will improve the uniformity of coverage across the sky). Redshift range is
from 0.015 to 0.1, $-48.3^{\circ}<\lambda<48.5^{\circ}$ and $6.25^{\circ}<\eta<36.25^{\circ}$
($\lambda$ and $\eta$ are the telescope coordinates). Our sample covers 2904
deg$^{2}$ on the sky. To ensure completeness, a volume-limited sample region was delineated. The final galaxy sample is approximately complete down to an absolute $r$-band magnitude limit -19.9 and contains 35,726 galaxies.

\begin{figure}[h]
\begin{center}
\includegraphics[scale=0.5]{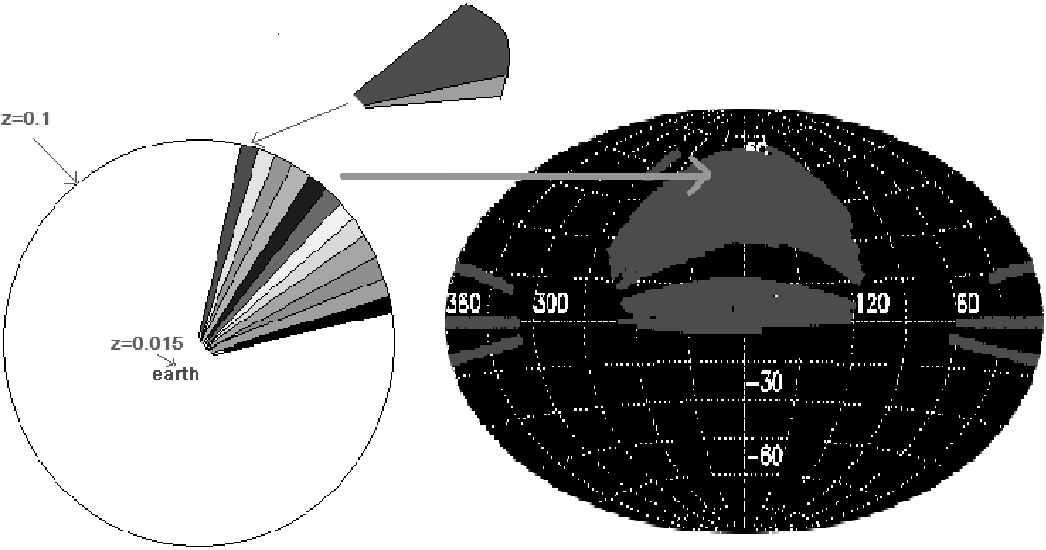}
\caption{Sketch of the sample geometry. The whole sample has been divided into 12 slices
($\thickapprox$ 2.5 degrees each) in $\eta$.} \label{geometry}
\end{center}
\end{figure}

We split the whole sample into 12 slices (see Fig. \ref {geometry}), which strictly follow the
survey coordinates ($\lambda, \eta$), each slice corresponding to a roughly east-west stripe on the sky. Fig. \ref{geometry} describes the observed sample geometry.                                                                      \vspace{10 mm}
\pagebreak
There are two major reasons for choosing 12 slices: (1) each slice is approximately two-dimensional
and (2) the slice-to-slice variations determine error bars, while keeping the number of objects per
slice at a fairly high level. Each slice includes around 2977 galaxies and those galaxy positions
are projected onto a two-dimensional image (projection perpendicular to the slice).

\begin{figure}[ht]
\centering
\includegraphics[scale=0.6,angle=270]{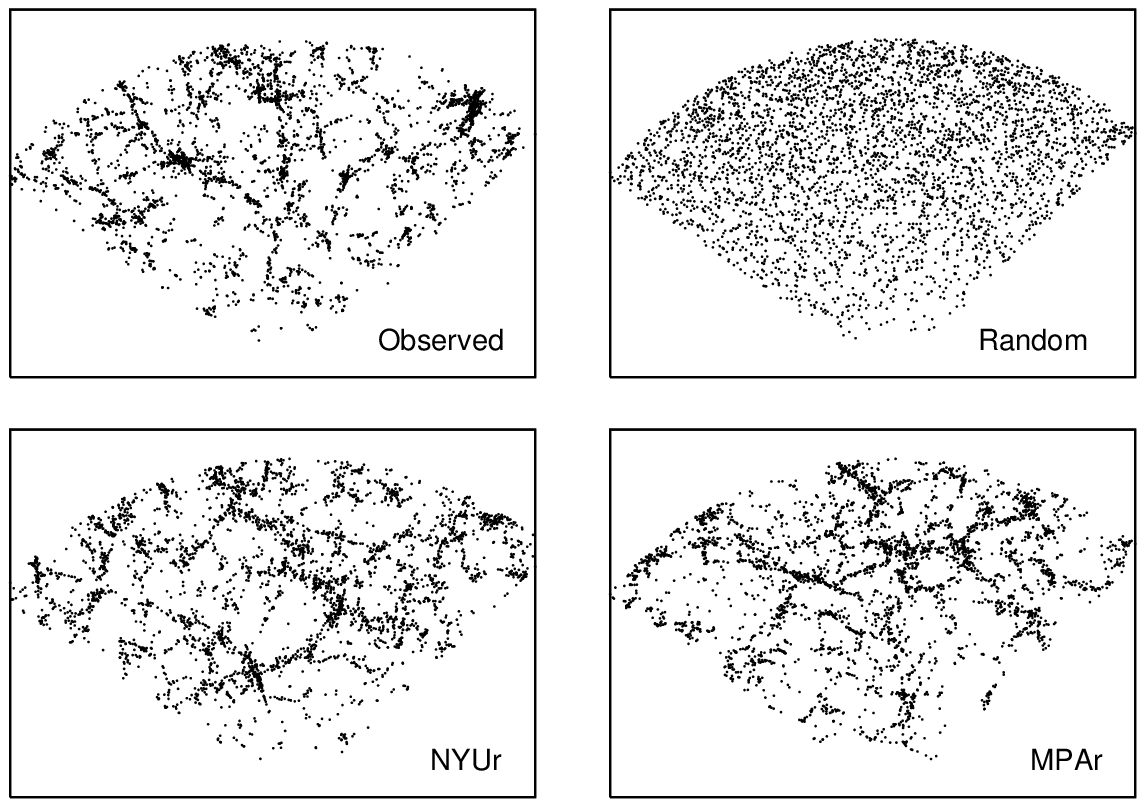}

  \caption{Example observed and mock samples slices.}
   \label{slice}
\end{figure}

Mock samples from two model universe simulations were used to compare with the observational
samples. One model universe is from the NYU Mock Galaxy Catalog \citep{ber06} in redshift
(velocity) space, henceforth referred as NYUr. They used the Hashed-Oct-Tree (HOT) code (Warren \&
Salmon 1993) to make \emph{N}-body simulations of the $\lambda$CDM cosmological model, with $\Omega_{m} =
0.3, \Omega_{\lambda} = 0.7, \Omega{b} = 0.04, h=H_{0}/(100 km s^{-1}Mpc^{-1})$ = 0.7, $n$ = 1.0, and
$\sigma_{8} = 0.9$. They identify
halos in the dark matter particle distributions using a friends-of-friends
algorithm with a linking length equal to 0.2 times the mean inter particle
separation. They then populate these halos with galaxies using a simple model
for the HOT of galaxies more luminous than a luminosity threshold. Every halo
with a mass $M$ greater than a minimum mass $M_{min}$ gets a central galaxy
that is placed at the halo center of mass and is given the mean halo
velocity. A number of satellite galaxies is then drawn from a Poisson
distribution with mean $<N_{sat}> = ((M \cdot M_{min})/M_{1})^{\alpha}$, for $M \geq
M_{min}$. These satellite galaxies are assigned the positions and velocities
of randomly selected dark matter particles within the halo. This model is in good agreement with a wide variety of cosmological observations (see, e.g., \citet{spe04,sel05,aba05}). Another mock sample is from the Millennium Run Semianalytic Galaxy Catalogue \citep{cro05} produced at the Max-Planck Institute for Astrophysics (MPA),
henceforth referred to as MPAr. The
simulation itself was carried out with a special version of the GADGET-2 code
\citep{spr01b, spr05b}. They use $\Omega_{m} =\Omega_{dm}+\Omega_{b}=0.3,
\Omega_{b} = 0.045,  h=0.73, n = 1.0$, and $\sigma_{8} = 0.9$. They apply in
post-processing an improved and extended version of the SUBFIND algorithm of
\citep{spr01a} to identify halos and a semianalytic model MODEL to build
galaxies. A third mock sample is an entirely randomly distributed set of points. Fig. \ref{slice} shows examples of slices from each sample used in this paper.

\pagebreak

\section{RESULTS AND DISCUSSION}
\begin{figure*}[!h]
  \centering
  \subfigure{
    \includegraphics[width=70mm,height=52mm]{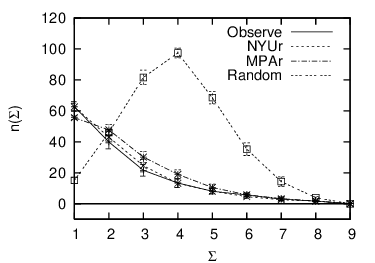}

  }
  \subfigure{
    \includegraphics[width=70mm,height=52mm]{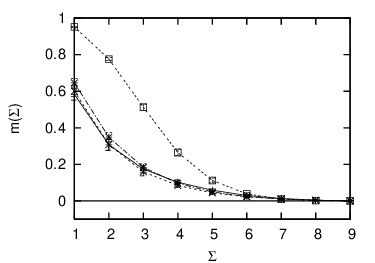}

  }

  \subfigure{
    \includegraphics[width=70mm,height=52mm]{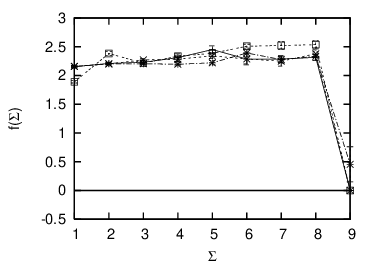}

  }
  \subfigure{
    \includegraphics[width=70mm,height=52mm]{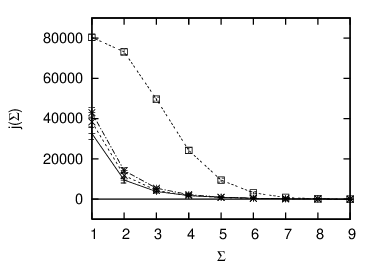}

  }
  \subfigure{
    \includegraphics[width=70mm,height=52mm]{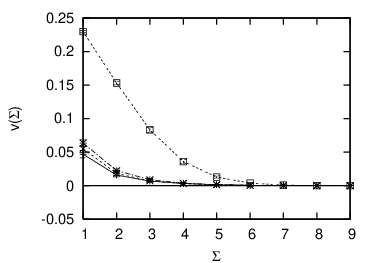}

  }

  \caption{Output functions from the MST for the size scale 15 Mpc:
Distribution of components ($n$), density ($m$), filament ($f$), pixels ($j$), and volume ($v$). For uniform
image, because there is only one value (the maximal pixel value equal with minimal value), no pixels
can be found above any threshold value, $n$=0, $m$=0, $f$=0, j=0, and $v$=0 (straight thick solid line)}
  \label{output_functions}
\end{figure*}

Fig. \ref{output_functions} shows the calculated output functions (Section \ref{sect_output}) for the
observational SDSS data, as well as for all mock sample data, where only the functions corresponding
to the size scale 15 Mpc (smoothed to that scale) are shown. The error bars are calculated from the
variance of the results over 12 slices for each sample (every sample has the same geometry for the
12 slices). The $x$-axis represents the threshold value $\Sigma$, which is linearly distributed
between the minimum ($\Sigma$=0) and maximum ($\Sigma$=10) pixel values for each smoothed slice.

\begin{figure}[ht]
\centering
\includegraphics[scale=0.7,angle=270]{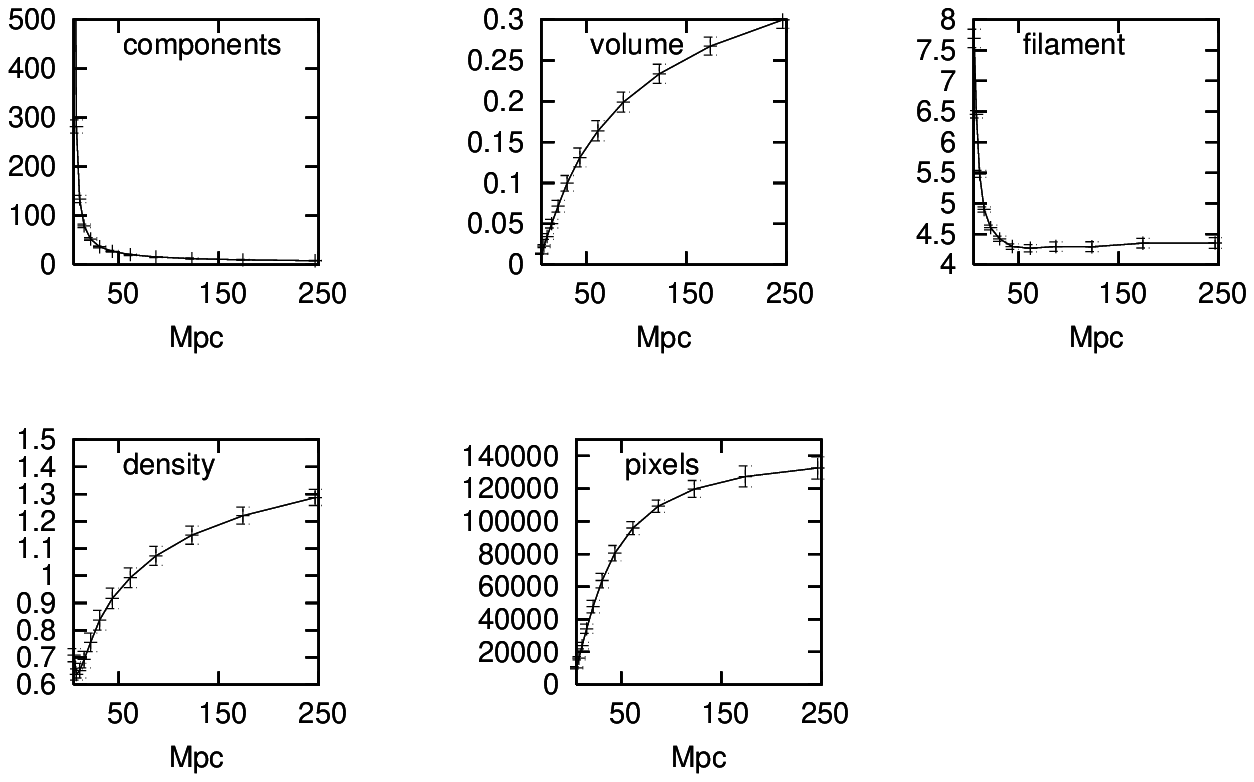}

  \caption{Results of the observed sample compared with uniform image on different filtering scales.
We calculate the distance between all samples and uniform images on every scale by Equation
(\ref{Euclidian}). The x-axis value is smoothing length and has ranges from 5 Mpc to 250 Mpc.}
   \label{ob_change}
\end{figure}

First we are interested to see  how far the observed sample is from the uniform image at each scale,
and we also want to see how different smoothing lengths influence the coordinates obtained from the
comparison. Fig. \ref{ob_change} displays the changes with smoothing scale. We find there is an
exponential change for all output functions.

To quantify the differences between all the mock sample curves and the observational curves such as
shown in Fig. \ref{output_functions}, Equation (\ref{Euclidian}) was used to calculate coordinates
in metric space. Each coordinate gives quantitative information as to ``how far'' the mock sample is
from the observed case. Coordinates were calculated for each output function, at each size scale.
Table 1 shows the coordinates, where, for simplicity, the scale sizes were categorized into four
groups (i.e., small, medium, large and huge scales). Also shown are the distance values obtained from
the maximal coordinate ranking scheme (Section \ref{sect_Euclidian}). Simply speaking, for each output
function at each scale group we find the maximal value first (among observed, NYUr, MPAr and random
samples), and then other values will be normalized by this maximal value (the maximal value itself
will be changed to ``1'' after normalization). In this way we normalize the different output
functions to sum them together. The resulting sum quantifies the overall differences between mock
sample data and observational data at each scale. The lower the distance value, the closer the mock
data is to the observed sample.
\begin{deluxetable}{p{0.8in}p{0.35in}p{0.35in}p{0.35in}p{0.35in}p{0.35in}p{0.35in}p{0.35in}}
  \tabletypesize{\scriptsize} \tablecaption{MST Coordinates and Overall
    Distance Between Models and Observational Data} \tablewidth{0pt} \tablehead{
    \colhead{ Filtering Scale} &
    \colhead{ Sample} &
    \colhead{ Components} &
    \colhead{ Density} &
    \colhead{ Filament} &
    \colhead{ Pixels} &
    \colhead{ Volume} &
    \colhead{ Maximal Ranking}
  }
 \startdata
Small   & NYUr  &   37.27   &   0.04    &   0.36    &   2611    &   0.003   &   0.75    \\
(5--10mpc)    & MPAr  &   62.99   &   0.03    &   0.41    &   3720    &   0.005   &   0.86    \\
    & Random    &   1230.98     &   0.38    &   0.69    &   48224   &   0.090   &   5   \\
\\
Medium  & NYUr  &   4.05    &   0.04    &   0.18    &   8021    &   0.012   &   0.62    \\
(15--30mpc)   & MPAr  &   12.78   &   0.11    &   0.34    &   13879   &   0.026   &   1.28    \\
    & Random    &   77.77   &   0.82    &   0.45    &   94207   &   0.284   &   5   \\
\\
Large   & NYUr  &   3.66    &   0.07    &   0.25    &   12174   &   0.024   &   0.78    \\
(40--80mpc)   & MPAr  &   5.71    &   0.17    &   0.33    &   17129   &   0.044   &   1.22    \\
    & Random    &   16.93   &   1.02    &   0.81    &   98539   &   0.341   &   5   \\
\\
Huge    & NYUr  &   2.57    &   0.10    &   0.33    &   17950   &   0.035   &   0.97    \\
(120--250mpc) & MPAr  &   2.81    &   0.24    &   0.47    &   24900   &   0.060   &   1.38    \\
    & Random    &   7.22    &   0.97    &   1.50    &   101196  &   0.320   &   5   \\

  \enddata

  \tablecomments{The maximal coordinate ranking used to calculate the distance
    takes only the new definition of the filament output function.}

\end{deluxetable}

\begin{figure*}[!h]
  \centering
  \subfigure{
    \includegraphics[width=70mm,height=56mm]{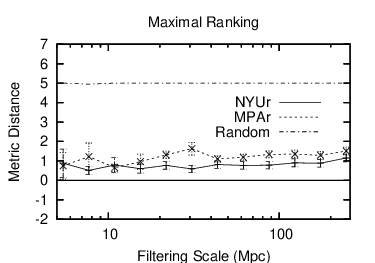}
    \label{fig:subfig1}
  }
  \subfigure{
    \includegraphics[width=70mm,height=56mm]{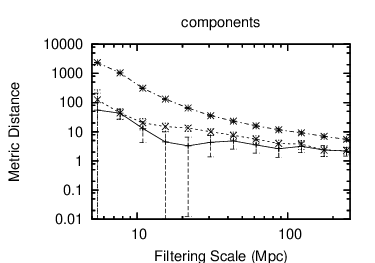}
    \label{fig:subfig2}
  }

  \subfigure{
    \includegraphics[width=70mm,height=56mm]{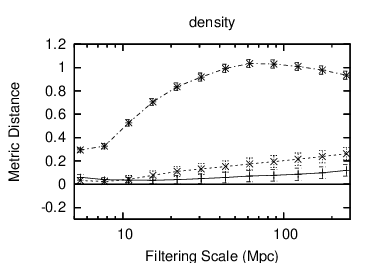}
    \label{fig:subfig3}
  }
  \subfigure{
    \includegraphics[width=70mm,height=56mm]{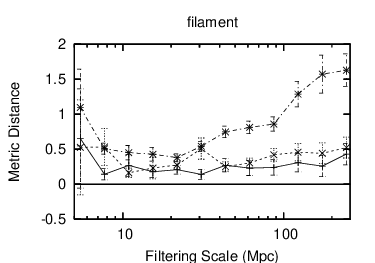}
    \label{fig:subfig4}
  }
  \subfigure{
    \includegraphics[width=70mm,height=56mm]{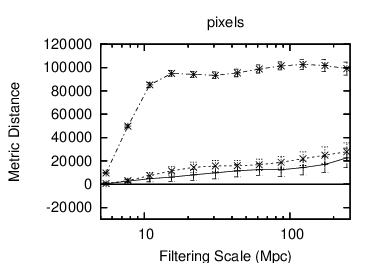}
    \label{fig:subfig5}
  }
 \subfigure{
    \includegraphics[width=70mm,height=56mm]{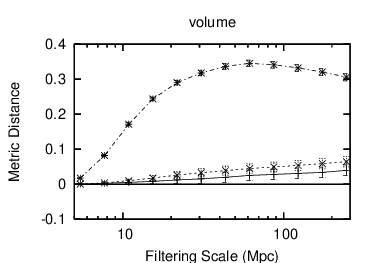}
    \label{fig:subfig6}
  }

  \caption{Metric distance (see Equation (\ref{Euclidian})) for maximal ranking result and output functions from mock samples.
  1$\sigma$ error bars are shown. We also plot the straight thick solid line representing
  the zero value of distance from the observed sample.}
  \label{distance_scale}
\end{figure*}

Table 1 clearly shows how the random mock sample is systematically the farthest from the
observational data. In order to get a better assessment of the more subtle differences from the
other mock samples, the distance values between each mock sample and the observational data were
plotted as a function of the size scale in Fig. \ref{distance_scale}, along with the rankings
obtained from the individual output functions.

\pagebreak

We can see that on small scales, both mock samples are within 1$\sigma$ error bar range for the
density, pixels and volume output function compared to the observations. We also note that the
random sample has a consistently large metric distance from the observed sample and that NYUr is
consistently and significantly closer to the observed sample case (zero values in Fig.
\ref{distance_scale}) than the MPAr simulation results. To investigate the reason for the difference
between the simulations, we repeated the above analysis, but using the mock samples in physical
space (MPAp and NYUp --- as opposed to redshift space MPAr and NYUr).

\begin{figure}[ht]
\centering
\includegraphics[scale=1.5]{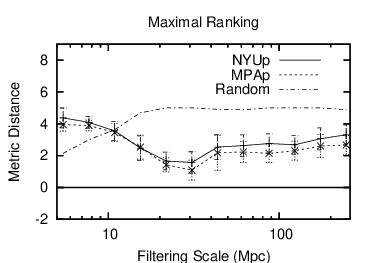}

  \caption{Metric distance (see Equation (\ref{Euclidian})) for maximal ranking result and output functions from mock samples
  in physical space.}
   \label{physical_space}
\end{figure}

While it is technically inappropriate to compare our redshift space observation sample with galaxy
distributions without velocity distortion, the results were informative. In Fig.
\ref{physical_space} we see on small scales that even the random sample is more closely matched to
the observed sample than NYUp and MPAp are. That is reasonable because the lack of redshift
distortion significantly changes the structure on small scales. We also see in Fig.
\ref{physical_space} that MPAp is closer to the observed sample statistics than NYUp in the maximal
ranking method (as well as most of the individual output functions). Considering the opposite
tendency for NYUr and MPAr, it is clear the two methods for assigning velocities to galaxies in the
MPA and NYU simulations make a noticeable difference.

\pagebreak

\section{CONCLUSION}

We have used a slightly modified MST of \citet{ada94} on multiple scale to study
the morphology of galaxy distributions. The technique gives a detailed morphological description of
galaxy distributions in metric space, on scales from about 5 Mpc to about 250 Mpc, with five output
functions showing strong statistical differences. We also find that the filament output function
values are high for the observations at small filtering scales but at around 50 Mpc the function
approaches a lower stable value. Considering that most voids in SDSS galaxies are around 30--50
Mpc \citep{gott05}, this seems a likely signature of those voids.

The key motivation for this work is to supplement traditional tools with a more informative way of quantifying the similarity in the ``visual'' morphological properties
between simulations and the observed universe. We use the ``metric distance'' as the parameter to describe that similarity through
multiple measures by calculating the value of each of the MST output functions. We combine the
values of each of the output functions into one ``final'' parameter for each simulation by the
maximal ranking method. In Table 1 and Fig. \ref{distance_scale}, it was demonstrated that two
\emph{N}-body simulations have done a similar job of approximating our universe and that NYUr is more close
to the observed sample than MPAr. From the analysis for Fig. \ref{physical_space}, we surprisingly
found that MPAp is more closely matched than the NYUp to the observed sample in redshift space, with
the implication that velocity determinations for simulation galaxies is a major contributor to the
relatively poorer match of the MPA simulation. The velocities of satellite galaxies in NYU
simulation halos are assigned randomly from the dark matter particles within the haloes
\citep{ber06}. However, in the MPA simulation, even satellite galaxies have interpolated velocities
(taken from the subhalo) rather than just randomly assigned ones  \citep{cro08}. It is very likely
that the mechanism for producing the velocity of satellite galaxies in MPA simulation contributes
noticeably to the relative shortcomings of MPAr in Fig. \ref{distance_scale}.

While the MST yields a single statistic for comparison of structure maps, in a way similar to other measures of large scale structure, we submit that its greater utility is in providing multiple intermediate outputs that convey insight into the physical differences between samples that lead to the statistical result. Of the many topological characteristics, threshold values, and scale samplings that the MST aggregates into a final result, we highlight a few examples of the specific physical differences that the technique reveals.

First, we have the expected result that the random sample is much different from all other samples at virtually all scales for all output functions. We have chosen only to use that case for normalization in the maximal ranking step.

Now, for our much more meaningful comparisons among MPA, NYU, and observed samples, we see that for the density output function, MPA has more high-density pixels  (at about the 5$\sigma$ level) than NYU sample, and the NYU sample has more high-density pixels than observed sample (1-2 $\sigma$). The volume and pixels output functions show similar trends as the density output function with the implication that those high-density pixels are also accompanied by large area regions of pixels above the various thresholds. The components and filament output functions are more complicated and fluctuate with the increasing scale. Simply speaking, for scales less than 50 Mpc, NYU and the observed sample are close to each other (around 1 $\sigma$), but MPA clearly has many more sizeable clumps (greater than 3$\sigma$) and is also more filamentary (greater than 3$\sigma$). For scales more than 50 Mpc, MPA shows more filamentary structure (greater than 3$\sigma$) than NYU sample, which is a little more filamentary (0.5-2 $\sigma$) than the observed sample. And observed sample has more clumps (1$\sigma$) than both mock samples.

Our next step is to apply the MST to the full three-dimensional galaxy distribution, which will require redevelopment/extension of output functions.

\acknowledgments The Millennium Run simulation used in this paper was carried out by the Virgo
Supercomputing Consortium at the Computing Center of the Max-Planck Society in Garching. The
semianalytic galaxy catalog is publicly available at
http://www.mpa-garching.mpg.de/galform/agnpaper. We thank Andreas A. Berlind for providing the NYU
Mock Galaxy Catalog.

\appendix

\section{Generalizing the Filament Index Definition}

\begin{figure}[h]
  \begin{center}
    \leavevmode
\includegraphics[scale=0.6]{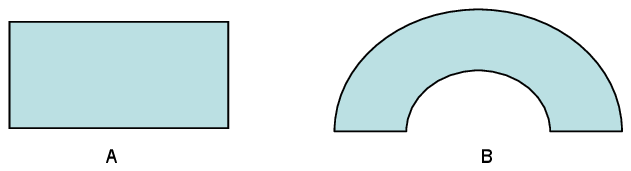}
    \caption{$A$ and $B$ have the same diameter and the same area and therefore, the same filament index,
    even though their structure is quite different.}
    \label{fault}
  \end{center}
\end{figure}

Let us first recall the definition of the filament index:
\begin{eqnarray}
F = {\pi D^2 \over {4 A}}.
\end{eqnarray}
Note first that $F$ depends only upon two values, $D$ and $A$, which are respectively the diameter
and the area of the component. Since we use the standard definition of a diameter (i.e., for a
component $S$, the diameter of $S$ is $D(S) = \max_{x,y \in S} \{|x-y|\}$), there is a possibility
that two components having quite different structures end up having the same filament index value
(Fig. \ref{fault}).
\vskip.1in \vskip.1in \vskip.1in \vskip.1in

The diameter, and therefore the filament index of non-convex components is under-estimated. Contrary
to what was originally said in \citet{kha04a}, the cause of this problem {\sl is not} the fact that
the definition of the diameter is not well adapted for non-convex
components. A closer look at the definition of the filament index shows that
one of its attributes is the circumference of a circle, $P_{\circ} = \pi D$:
\begin{eqnarray}
F = {\pi D^2 \over {4 A}} = {(\pi D) D \over {4 A}} = {P_{\circ} D \over {4 A}}.
\end{eqnarray}
So by definition, the filament index ``expects'' to be treating convex objects (a circle certainly
being the most trivial example of a convex object). And that is where the change should be made:
Instead of changing the definition of the diameter, one should simply change the definition of the
perimeter to have it in its most general form, $P$. So the generalized version
of the filament index is therefore
\begin{eqnarray}
F = {P D \over {4 A}},
\end{eqnarray}
where $P$ is the perimeter of the underlying object.
\vskip.1in \vskip.1in \vskip.1in \vskip.1in

\begin{figure}[h]
  \begin{center}
    \leavevmode
    \includegraphics[scale=0.8]{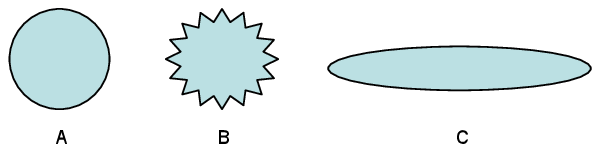}
    \caption{Objects $A$, $B$, and $C$ in order of increasing filament index value. All have the same
    area. $A$ and $B$ have the same diameter, but since the perimeter of $B$ is larger, their (newly
    generalized) filament index is different.  Since object $C$ has a larger diameter and a larger
    perimeter than object $A$, it therefore has a larger filament index. And although objects $B$ and
    $C$ have the same perimeter, since object $C$ has a larger diameter, it has a larger filament index.}
    \label{3blobs}
  \end{center}
\end{figure}

 \vskip.1in

One can readily see from Fig. \ref{fault} that although both objects have the same diameter and
area, since their perimeter is quite different, object B will have the larger filament index, which
is what one would intuitively expect. In Fig. \ref{3blobs} are shown three objects of increasing
filament index value. One can easily see how the newly generalized filament index definition will
greatly help in the distinction of different geometrical features in the analyzed objects (or
components). However, this new definition of $F$ is still degenerate in the sense that one can still
find an infinite number of objects having the same $F$. However, the degree of degeneracy is much
less than for the standard, original definition of $F$.


\begin{thebibliography}{}


\bibitem[Abazajian(2005)]{aba05} Abazajian, K., Kadota, K., \& Stewart, E. D.

    2005, JCAP, 0508, 008

\bibitem[Adams(1992)]{ada92} Adams, F. C. 1992, \apj, 387, 572

\bibitem[Adams \& Wiseman(1994)]{ada94} Adams, F. C. \& Wiseman, J. 1994,
    \apj, 435, 693

\bibitem[Adelman-McCarthy et al.(2007)]{jen07} Adelman-McCarthy, J. K., et
    al. 2007, \apjs, 172, 634

\bibitem[Berlind et al.(2006)]{ber06} Berlind, A. A., et al. 2006, \apjs,
    167, 1

\bibitem[Blanton et al.(2005)]{bla05} Blanton, M. R., et al. 2005, \aj, 129,

    2562

\bibitem[Coles \& Lucchin(1995)]{cole95} Coles, P. \& Lucchin, F. 1995,
    Cosmology: The Formation and Evolution of Cosmic
    Structure (NY:Wiley)

\bibitem[Crotonet al.(2005)]{cro05} Croton, D. J., et al. 2006, \mnras, 365,
    11

\bibitem[Croton(2008), private communication]{cro08} Croton, D. J. 2008, private communication.

\bibitem[Dolag et al.(2008)]{dol08} Dolag, K., Borgani, S., Schindler, S., Diaferio, A., \&
Bykov, A. M. Space Sci.Rev., 134, 229

\bibitem[Donoho(1988)]{don88} Donoho, D. L. 1988, Ann. Stat., 16, 1390

\bibitem[Gill et al.(2004)]{gill04}Gill, S. P. D., Knebe, A., \& Gibson, B.
    K. 2004, \mnras, 351, 399

 \bibitem[Gott et al.(2005)]{gott05} Gott, J. R., Juri\'{c}, M., Schlegel, D.,
    Hoyle, F., Vogeley, M., Tegmark, M., Bahcall, N., \& Brinkmann, J. 2005, \apj 624, 463

\bibitem[Khalil(2004)]{kha04b} Khalil, A. 2004, PhD thesis, Univ. Laval (available online at
    http://www.theses.ulaval.ca/2004/22165/22165.pdf)

\bibitem[Khalil et al.(2004)]{kha04a} Khalil, A., Joncas, G., \& Nekka, F.
    2004, \apj, 601, 352

\bibitem[Khalil et al.(2006)]{kha06} Khalil, A., Joncas, G., Nekka, F.,
  Kestener, P., \& Arneodo, A. 2006, \apjs, 165, 512

\bibitem[Kiang, Wu \& Zhu(2004)]{wu04} Kiang, T., Wu, Y., \& Zhu, X. 2004,
    Chin. J. Astron. Astrophys., 3, 209

\bibitem[Martinez et al.(2005)]{vic05}Martinez, V. J., Starck, J. L., Saar, E.,
    Donoho, D. L., Reynolds, S., Cruz .P., \&
    Paredes, S. 2005, \apj, 634, 744

\bibitem[Peebles(1980)]{peeb80}Peebles, P.J.E. 1980, Principles of Physical
    Cosmology, Princeton University Press

\bibitem[Saar et al.(2007)]{saar07}Saar E., Martinez V.J., Starck J.-L., Donoho D.L. 2007, \mnras, 374, 1030

\bibitem[Sahni et al.(1997)]{sah97} Sahni, V., Sathyaprakash, B.S., \& Shandarin,
    S.F. 1997, \apj, 476, L1

\bibitem[Seljak et al.(2005)]{sel05} Seljak, U., et al. 2005, \prd 71,
    103515

\bibitem[Shandarin(1983)]{sha83} Shandarin, S.F. 1983, Soviet Astron. Lett.,

    9, 104

\bibitem[Silverman(1981)]{sil81} Silverman, B. W. 1981, J. R. Stat. Soc.
    B, 43, 97

\bibitem[Skrutskie et al.(2000)]{skr00} Skrutskie, M. F., et al. 2006, \aj,
    131, 1163

\bibitem[Spergel et al.(2004)]{spe04} Spergel, D. N., et al. 2004, \apj, 608, 10

\bibitem[Springel et al.(2005)]{spr05} Springel, D. N., et al. 2005, \nat,
    435, 629

\bibitem[Springel et al.(2005)]{spr05b} Springel, D. N., et al. 2005, \mnras,

    364, 1105

\bibitem[Springel et al.(2001a)]{spr01a} Springel V., White S. D. M., Tormen

    G., Kauffmann G., 2001a, \mnras, 328, 726

\bibitem[Springel et al.(2001b)]{spr01b} Springel V., Yoshida N., White S. D.

    M., 2001b, New Astron., 6, 79

\bibitem[Taylor et al.(2003)]{tay03} Taylor, A. R., et al. 2003, \aj, 125,
    3145

\bibitem[Warren \& Salmon(1993)]{war93} Warren M.S., Salmon J.K., A parallel hashed oct-tree N-body algorithm. In Supercomputing '93, pages 12-21, Los Alamitos, 1993. IEEE Comp. Soc.

\bibitem[Weinberg(2005)]{wei05}Weinberg, D. H. 2005, Science, 309, 564

\bibitem[White(1979)]{white79} White S.D.M. 1979, \mnras 186,145

\bibitem[Wiseman \& Adams(1994)]{wis94} Wiseman, J. \& Adams, F.C. 1994,
    \apj, 435, 708

\bibitem[Wu, Batuski \& Khalil(2008)]{wu08} Wu, Y., Batuski, D. J., \& Khalil,
    A. 2008, The Fractal Structure of the
    Universe (Germany
:VDM Verlag Dr. Mueller
e.K.)

\bibitem[York et al.(2000)]{york00} York, D. G., et al. 2000, \aj, 120, 1579

\bibitem[Zeldovich(1982)]{zel82a} Zeldovich, Ya. B. 1982, Soviet Astron.
    Lett., 8, 102


\end{thebibliography}
\end{document}